\documentclass[prb,twocolumn,superscriptaddress,floatfix,noshowpacs,10pt,longbibliography]{revtex4-2}%

\usepackage{graphicx,bm,times}
\usepackage{amsmath}
\usepackage{amsfonts}
\usepackage{amssymb}

\usepackage{bm}
\usepackage{color}
\usepackage{times}

\usepackage[%
colorlinks=true,
urlcolor=blue,
linkcolor=blue,
citecolor=blue
]{hyperref}

\begin{document}

\title{Multi-band description of the upper critical field of bulk FeSe}

\author{M. Bristow}
\affiliation{Clarendon Laboratory, Department of Physics,
University of Oxford, Parks Road, Oxford OX1 3PU, UK}

\author{A. Gower}
\email{Current address: Cavendish Laboratory, University of Cambridge,
 Cambridge CB3 0HE, United Kingdom}
\affiliation{Clarendon Laboratory, Department of Physics,
University of Oxford, Parks Road, Oxford OX1 3PU, UK}

\author{J. C. A. Prentice}
\affiliation{Department of Materials, University of Oxford, Parks Road, Oxford OX1 3PH, United Kingdom}

\author{M. D. Watson}
\email{Current address: Diamond Light Source, Division of Science, Didcot, OX11 0DE, UK} \affiliation{Clarendon Laboratory, Department of Physics,
	University of Oxford, Parks Road, Oxford OX1 3PU, UK}

\author{Z. Zajicek}
\affiliation{Clarendon Laboratory, Department of Physics,
	University of Oxford, Parks Road, Oxford OX1 3PU, UK}

\author{S. J. Blundell}
\affiliation{Clarendon Laboratory, Department of Physics,
	University of Oxford, Parks Road, Oxford OX1 3PU, UK}

\author{A. A.\;Haghighirad}
\affiliation{Clarendon Laboratory, Department of Physics,
University of Oxford, Parks Road, Oxford OX1 3PU, UK}
\affiliation{Institute for Quantum Materials and Technologies, Karlsruhe Institute of Technology, 76021 Karlsruhe, Germany}

\author{A. McCollam}
\email{Current address: School of Physics, University College Cork, Cork, Ireland}
\affiliation{High Field Magnet Laboratory (HFML-EMFL), Radboud University, 6525 ED Nijmegen, The Netherlands}

\author{A. I. Coldea}
\email[corresponding author: ]{amalia.coldea@physics.ox.ac.uk}
\affiliation{Clarendon Laboratory, Department of Physics,
University of Oxford, Parks Road, Oxford OX1 3PU, UK}

\begin{abstract}
The upper critical field of multi-band superconductors can be an essential quantity to unravel the nature of superconducting pairing and its interplay with the electronic structure.
Here we experimentally map out the complete upper critical field phase diagram of FeSe for different magnetic field orientations at temperatures down to $0.3\,$K using both resistivity and torque measurements.
The temperature dependence of the upper critical field reflects that of a multi-band superconductor and requires a two-band description in the clean limit with band coupling parameters favouring interband over intraband interactions.
Despite the relatively small Maki parameter in FeSe of $\alpha\sim1.6$, the multi-band description of the upper critical field is consistent with the stabilization of a
FFLO state below $T/T_{\rm{c}}\sim 0.3$.
We find that the anomalous behaviour of the upper critical field is linked to a departure from the single-band picture,
and FeSe provides a clear example where multi-band effects and the strong anisotropy of the superconducting gap need to be taken into account.
\end{abstract}
\date{\today}
\maketitle

\section{Introduction}

The upper critical field of the iron-based superconductors have  unusually large values and reveal
the interplay of orbital and paramagnetic pair-breaking effects in multi-band superconductors with competing pairing channels and unconventional pairing symmetry \cite{Hirschfeld2011}.
These systems can provide ideal conditions to stabilize exotic superconducting phases like the Fulde-Ferrell-Larkin-Ovchinnikov (FFLO) state \cite{Gurevich2010,Song2019}, in which the order parameter varies in space.
This exotic superconducting phase can be stabilized in multi-band clean materials due to the likely presence of shallow bands \cite{Mou2016} and the consequent large Pauli paramagnetic effects \cite{Gurevich2010}.
Furthermore, iron-based superconductors have complex electronic structures formed of electron and hole like bands with different orbital character and highly anisotropic superconducting gaps.
These effects are likely to manifest in the behaviour of the temperature dependence of the upper critical field.

FeSe is a clean iron-based superconductor that offers us a platform on which to perform a detailed study of the upper critical fields down to the lowest temperature as it has a relatively low superconducting transition temperature, $T_{\rm{c}}$, of $\sim9\,$K.
FeSe exhibits a nematic electronic state below $\sim90\,$K, which leads to a small and strongly anisotropic Fermi surface driven by orbital degrees of freedom and interactions that trigger an orthorhombic-tetragonal structural transition in the absence of the long-range magnetic order \cite{Coldea2018,Watson2015a,Watson2017a,Coldea2021}.
The same interactions that affect the electronic structure are likely to influence superconductivity, leading to highly anisotropic superconducting gaps and potential orbital-dependent pairing interactions \cite{Hanaguri2018,Sato2018,Sprau2017}.
FeSe, due to the presence of its shallow bands, has been proposed to be a possible candidate for the BCS-BEC crossover \cite{Kasahara2014,Kasahara2016}, where the superconducting gap size and the superconducting transition temperature $T_{\rm{c}}$, are comparable to the Fermi energy.
However, no evidence of a pseudogap of preformed pairs above $T_{\rm{c}}$ has been detected \cite{Hanaguri2019}.
Instead, the multi-band nature of the system together with large anisotropies that develop inside the nematic state are likely to govern many of the features of this exotic superconductor.
Recently, it was proposed that   FeSe
has unusual field-induced phases
from a Fulde-Ferrel-Larkin-Ovchinnikov (FFLO) state,
where the formation of planar nodes gives rise
to a segmentation of the flux-line lattice \cite{Kasahara2020,Kasahara2021},
or a field-induced magnetic phase transition in-plane magnetic field \cite{Jong2020}.

In this paper we study the upper critical fields of FeSe as a function of temperature down to $\sim0.33\,$K in magnetic fields
up to $35\,$T for two magnetic field orientations with respect to the conducting planes using electrical transport and magnetic torque measurements.
We find that the upper critical field of FeSe shows a clear deviation from a single-band  description,
with a strikingly linear behaviour down to the lowest temperatures with the magnetic field perpendicular to the conducting planes.
Furthermore, we find that a two-band model can describe the temperature dependence of the upper critical field for both orientations.
The band coupling parameters indicate the dominance of the interband pairing channels.
At low temperatures, below $\sim2\,$K, when the magnetic field is aligned parallel to the conducting planes, the
upper critical field of FeSe does not saturate but displays a characteristic upturn,
consistent with the emergence of a FFLO state
that can be stabilized in the presence of Pauli-paramagnetic effects.

\begin{figure*}[htbp]
	\centering
	\includegraphics[trim={0cm 0cm 0cm 0cm}, width=1\linewidth,clip=true]{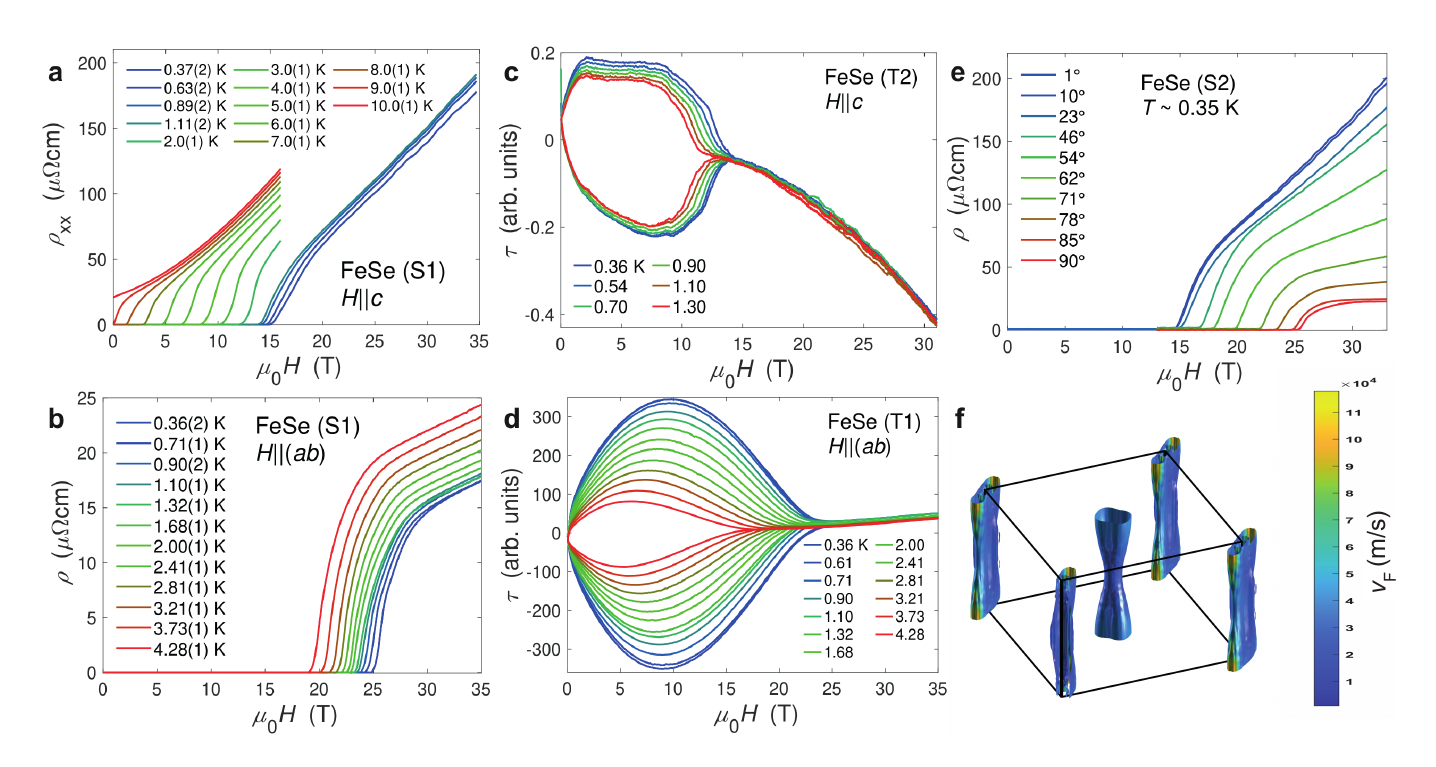}
	\caption{
		(a) and (b) Resistivity versus applied magnetic field for FeSe (S1) at different constant temperature for $H||c$ and $H||(ab)$ respectively.
		(c) and (d) Torque versus magnetic field at constant temperatures for $H||c$ (T2) and $H||(ab)$ (T1) respectively.
		The torque has arbitrary units.
		(e) Resistivity against applied magnetic field for FeSe (S2) at $\sim0.35\,$K at different angles in magnetic field.
		(f) Calculated Fermi surface of FeSe with experimental lattice parameters, shifted to match quantum oscillations data
 (as detailed in Fig.~S3
  in the SM) \cite{Terashima2014,Watson2015a,Coldea2019}.
The colours in (f) reflect the variation of the Fermi velocity from 200 meV \AA~ for in-plane  electron pocket  to 23 meV \AA~ for its related out-of-plane value.
The corresponding anisotropy is $\Gamma$=$\lambda_{c}/\lambda_{ab}\sim$  4.5.
}
	\label{Fig1}
\end{figure*}

\section{Experimental Methods}

Single crystals of FeSe were grown by the KCl/AlCl$_3$ chemical vapour transport method \cite{Bohmer2013} and were from the same batches reported previously in Ref.~\cite{Watson2015a,Bristow2020}.
Electrical transport was measured using the low-frequency 4- or 5-point contact method.
Low resistance contacts ($<1\,\rm{\Omega}$) were made using indium solder to reduce heating at low temperatures.
Typically a current of $1\,$mA was used for electrical transport.
Quantum oscillations are observed in these samples, indicating high-quality single crystals, as described in Refs.~\cite{Watson2015a,Bristow2020}.
Torque measurements were performed using two cantilevers, one with a sample and another dummy lever in a Wheatstone bridge configuration and small ac currents of $\sim70\,\mu$A at a low frequency ($\sim 72\,$Hz).
Low magnetic field measurements (up to $16\,$T) were performed using a Quantum Design Physical Properties Measurements System (PPMS).
Both field sweeps at fixed temperature and temperature sweeps in constant field were used to build the upper critical field phase diagrams.
High field measurements were performed at the High Field Magnetic
Laboratory, (HMFL-EMFL) Nijmegen up to $38\,$T at temperatures down to $\sim0.3\,$K.
The upper critical fields were identified from magnetic field sweeps at constant
temperature and the angular dependence was measured using a single-axis rotator.

\section{Results and Discussion}

{\bf Transport and torque measurements.}
Fig.~\ref{Fig1}(a)-(d) show resistivity (S1) and torque data (T1 and T2) for FeSe against magnetic field for different sample orientations in relation to the applied magnetic field.
The superconducting transition temperature
for the transport sample S1 is $T_{\rm c} \sim 9.08$~K, which is defined by the offset temperature.
As the magnetic field increases the superconducting transition becomes broader and is suppressed at $\sim14.6(1)\,$T for $H||c$ at $0.36\,$K  (Fig.~\ref{Fig1}(a))
whereas  the value is above $25\,$T for $H||(ab)$, as the orbital effects are less effective in destroying superconductivity in this orientation.
These values are in good agreement with previous reports \cite{Kasahara2020,Jong2020}.
Furthermore, significant broadening of the superconducting transition for $H||(ab)$
was observed in other systems, such as for CaKFe$_4$As$_4$  \cite{Bristow2020CaK}
and FeSe$_{0.5}$Te$_{0.5}$ \cite{Serafin2010}, where it was attributed to strong anisotropic superconducting fluctuations .
In our transport data, the upper critical field, $\mu_{0}H_{\rm{c2}}$, is defined as the offset magnetic field, as shown in
 Fig.~S1
in the Supplementary Material (SM).
Additionally, we use torque magnetometry, as a thermodynamic probe,
to detect the irreversible magnetic field, $H_{\rm{irr}}$,
for additional single crystals (T2 and T3) for different magnetic field orientations,
as shown in Fig.~\ref{Fig1}(d) and
Fig.~S1
in the SM.

\begin{figure}[htbp]
	\centering
\includegraphics[trim={0cm 0cm 0cm 0cm}, width=0.90\linewidth,clip=true]{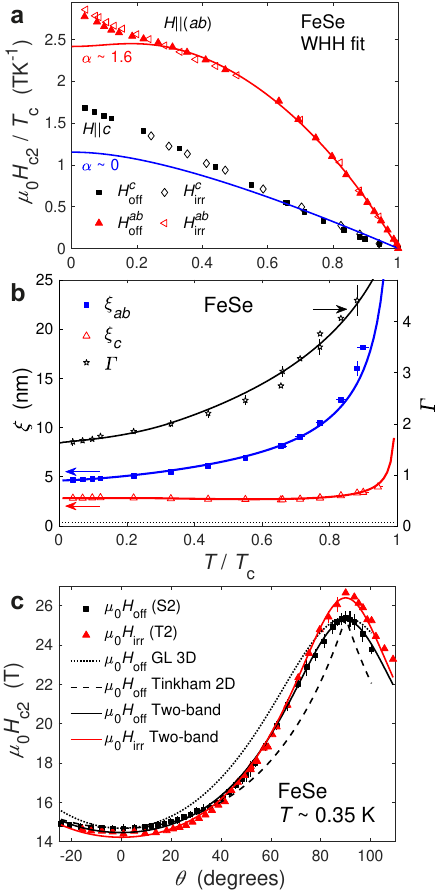}
	\caption{
		(a) Upper critical fields of FeSe for $H||c$ and $H||(ab)$ obtained from transport (solid symbols) and
torque measurements (open symbols) scaled against reduced temperature, $T/T_{\rm{c}}$,
where the superconducting transition temperature, $T_{\rm{c}}$,  is 9.08 and 8.5~K, respectively.
		The blue and red lines are WHH fits for  $H||c$ and $H||(ab)$ respectively, using different $\alpha $ values.
		(b) The temperature dependence of coherence lengths (left-axis, blue and red) and anisotropy of the upper critical field (right-axis, black).
		The $H_{\rm{c2}}(T\rightarrow0)$ values were used to find the zero-temperature coherence lengths.
		The horizontal dashed line represents the 3D-2D crossover length of $\sim{}c/\sqrt{2}$, where $c$ is the $c$-axis lattice constant.
		Solid lines are guide to the eye using the two-band model (described later).
		(c) Angular dependence of the superconducting upper critical field, $\mu_{0}H_{\rm{c2}}(\theta)$, at $0.35\,$K for different single crystals of FeSe
 (raw data are shown in
 Fig.~S2
 in the SM).
		The dotted black line is a fit to the 3D Ginzburg-Landau (GL) model and the dashed black line represents a fit to the 2D
Tinkham model \cite{Tinkham1996}, both for upper critical fields from resistivity data.
		The solid black and red lines are two-band fits to upper critical fields from resistivity and torque respectively.
	}
	\label{Fig2}
\end{figure}

{\bf Single-band description.}
From the transport and torque data we have constructed a complete upper critical field phase diagram of FeSe down to $\sim0.35\,$K in magnetic fields up to $35\,$T.
This is shown in reduced units for the two orientations in Fig.~\ref{Fig2}(a).
In order to assess the role of orbital and Pauli paramagnetic effects on the upper critical field of FeSe we first describe the temperature
dependence using the three-dimensional Werthamer-Helfand-Hohenberg (WHH) model for a single-band weakly-coupled superconductor with an ellipsoidal Fermi surface \cite{Werthamer1966,Helfand1966}.
The slope close to $T_{\rm{c}}$ ($H'_{\rm{c2}}=-|\mu_0 dH_{\rm{c2}}/dT|_{T_{\rm{c}}}$) can be used to
estimate the zero-temperature orbital upper critical field using $\mu_0 H_{\rm{orb}}(0)=0.73T_{\rm{c}}H'_{\rm{c2}}$ in the clean limit \cite{Werthamer1966}.
This gives $\sim11.1\,$T for $H||c$ (using a slope of $-1.7(1)\,$T/K  and $T_{\rm{c}}\sim9\,$K), smaller than the measured value
of $\sim14.6\,$T at $0.3\,$K, whereas   $\mu_0$ $H_{\rm{orb}}(0)\sim42.7\,$T for $H||(ab)$ (using a slope of $\mu_0 H'_{\rm c2}\sim-6.5(1)\,$T/K).
Furthermore, the upper critical field of FeSe for $H||c$ in Fig.~\ref{Fig2}(a) has a striking linear dependence, similar to other reports on FeSe \cite{Kasahara2016,Kasahara2020}.
This has also been observed in LiFeAs \cite{Khim2011}.
This is a significant deviation from single-band behaviour (Fig.~\ref{Fig2}(a) blue and red lines), and suggests that a multi-band model needs to be considered \cite{Gurevich2003}.

Besides the orbital-limiting effects, the BCS Pauli paramagnetic limit, determined by the magnetic field at which Zeeman splitting breaks the spin-singlet Cooper pair, is defined for a single gap in the weak coupling limit as $ \mu_0H^{\rm{BCS}}_{\rm{P}}(0)=1.85T_{\rm{c}}$ \cite{Clogston1962} and reaches a value of $16.6\,$T.
However, FeSe is not a single gap superconductor and the Pauli limit can be estimated using the value of the superconducting gap
with $\mu_0 H^{\Delta_{i}}_{\rm{P}}(0)=\sqrt{2}\Delta_{i}/g\mu_{\rm{B}}$,
where $\mu_B$ is the Bohr magneton, and $g$ is the Land\'{e} factor \cite{Werthamer1966}.
Measured values for the superconducting gaps, $\Delta_{i}$, vary  between $2.3\,$meV for the hole pocket near the $\Gamma$ point to $1.5\,$meV for the electron
pocket at the M point \cite{Sprau2017}, with another potential small gap of $0.39\,$meV suggested from specific heat data \cite{Sun2017}
(values of 1.8~meV and 0.7~meV are also extracted from superfluid density \cite{Biswas2018}).
Using these values, the Pauli paramagnetic field  $\mu_0 H^{\Delta_i}_{\rm P}(0)$
would vary between $28\,$T, $18.3\,$T and $4.8\,$T, respectively (assuming $g=2$).
The Pauli limiting field can also be enhanced by either strong coupling effects (from electron-boson coupling or correlations)
or significant spin-orbit scattering
(where finite $\lambda_{\rm{SO}}$ would reduce $g$ below 2).
Thus, Pauli limiting could exceed the single-band estimate, since FeSe is a multi-band system with several
anisotropic superconducting gaps \cite{Sprau2017} and the largest gap is expected to determine
 the Pauli limit \cite{Gurevich2010,Gurevich2014}.

The orbital pair breaking dominates the temperature dependence for $H||c$,
only down to $0.6 T_{\rm{c}}$ and afterwards it deviates significantly, as shown in Fig.~\ref{Fig2}(a). 	
Similarly, for $H||(ab)$ the deviation from the single-band WHH model is obvious below $0.25T_{\rm{c}}$.
The parameter used to quantify the Pauli pair breaking contribution to the upper critical field, when the magnetic field is aligned along the conducting $(ab)$-plane,
is the Maki parameter, $\alpha$=$\sqrt{2} H_{\rm{orb}}(0)/H_{\rm{P}}(0)$.
This varies strongly depending on the Pauli paramagnetic limiting field considered, with values ranging
from $\alpha \sim 3.6$,  in the BCS limit, towards $\sim 2.15-3.3$ using $H^{\Delta_{1,2}}_{\rm{P}}(0)$.
Using the measured lowest temperature value of $\mu_0 H_{\rm{c2}}(0)\sim 26$~T,
the standard form $\mu_{0}H_{\rm{c2}}(0)=\mu_{0}H^{\rm{orb}}(0)/\sqrt{1+\alpha^2}$
would give $\alpha \sim 1.3$.
A more reliable method of extracting the $\alpha$ value is by fitting $H_{\rm{c2}}(T)$ to the WHH model which
gives $\alpha=1.6$ for $H||(ab)$, as shown in Fig.~\ref{Fig2} for $H||c$ (solid blue line)  and $H||(ab)$ (red line), respectively.
In estimating the value of $\alpha$  the spin-orbit constant $\lambda_{\rm{SO}}$ was close to zero,
different from those found for thin flakes of FeSe, where $\lambda_{\rm{SO}}\sim0.2$  and $\alpha\sim 2.4$ \cite{Farrar2020}.
Interestingly, a relatively low value of $\alpha$ would also imply the important role played by the largest gap on the
hole band dominating the Pauli paramagnetic effects. 
For a clean isotropic single-band system, the Maki parameter is given by $\alpha=\pi^{2}\Delta/(4 E_{\rm{F}})$, and
thus one can estimate the Fermi energy as $\sim 3.5\,$meV, which would suggest
the presence of a shallow band, of a similar size to the smallest pocket observed in quantum oscillations \cite{Terashima2014,Watson2015a}.
The $\alpha$ value for FeSe is similar to other systems, such as LiFeAs \cite{Khim2011}, but much smaller
than that of FeTe$_{0.6}$Se$_{0.4}$ single crystals where $\alpha\sim5.5$ \cite{Khim2010} or
CaKFe$_4$As$_4$ with $\alpha\sim4.2 $ \cite{Bristow2020CaK}, where the upper critical field is strongly dominated by Pauli paramagnetic effects.
Strong paramagnetic effects seem to be an important signature of optimally doped iron-based superconductors \cite{Bristow2020CaK,Tarantini2011,Cho2011,Gurevich2011a,Fuchs2009}.
It is worth emphasising that in the related isoelectronic compounds Fe$_{1+y}$Se$_{x}$Te$_{1-x}$ ($x=0.4$),
where BEC-BCS crossover has been invoked to be present due to a hole band very close to the Fermi level \cite{Rinott2017},
the upper critical field studies have found much larger values of $\alpha$ than those extracted here for FeSe.

Fig.~\ref{Fig2}(b) shows that  the anisotropy of the upper critical field of FeSe,
defined as the ratio of the upper critical field for different orientations,  $\Gamma=H_{\rm{c2}}^{ab}/H_{\rm{c2}}^{c}$, drops dramatically with decreasing temperature from $\sim 4$ to $\sim 1.7$,
in good agreement with heat capacity data \cite{Hardy2020} and similar to other iron-based superconductors like LiFeAs \cite{Khim2011}.
The Fermi surface of FeSe has significant warping on the outer electron and hole bands that can potentially allow circulating currents out
of plane (see Fig.~\ref{Fig1}(f)).
Furthermore, the calculated anisotropy of the penetration depth based on plasma frequencies (as in Ref.~\onlinecite{Hashimoto2010})
 is $\Gamma$=$\lambda_{c}/\lambda_{ab}\sim$  4.5  (see
 Fig.~S3
 in the SM for the shifted Fermi surface).
 This is similar to the measured anisotropy of $\sim 4$ close to $T_{\rm c}$ (Fig.~\ref{Fig2}(b)),
 suggesting that the Fermi surface details play an important role in understanding its superconducting properties.
The upper critical fields for the two orientations do not cross at low temperatures as found in optimally doped iron-based superconductors,
where Pauli paramagnetic effects are significant, such as FeSe$_{0.5}$Te$_{0.5}$ \cite{Serafin2010,Braithwaite2010}, and CaKFe$_4$As$_4$ \cite{Bristow2020CaK}.

Using the experimental values of the upper critical fields for different orientations in magnetic field, we can extract the coherence lengths for FeSe in different temperature regimes, as shown in Fig.~\ref{Fig2}(b).
The coherence lengths at low temperature reach values of $\xi_{ab}=4.57(4)\,$nm and $\xi_{c}=2.78(6)\,$nm for FeSe and are well above the $c/\sqrt{2}\sim0.388\,$nm limit, where a 3D-2D crossover can occur, as was observed in the thinnest flakes of FeSe \cite{Farrar2020}.
The mean free path due to elastic scattering from impurities of FeSe was found to vary between $\ell \sim  712$~\AA~ \cite{Zajicek2022} and 850~nm \cite{Kasahara2020}
using the $\ell$-isotropic approximation. This approach assumes that at the lowest temperatures, where the transport properties are dominated
 by large angle scattering from impurities (rather than to the quantum scattering time \cite{Farrar2022}), the in-plane mean-free path is limited only by the separation between the scattering impurities and
  it is not sensitive to the Fermi velocity across the Fermi surface or on different sheets in the case of a multiband metal \cite{Hussey1998}.
  Thus, the mean free path only depends on the averaged Fermi surface formed of two-dimensional cylinders with compensated hole and two electron pockets, approximated by $k_F \sim 0.1$~\AA$^{-1}$.
 These values are much larger than the coherence length, $\xi(0)\ll\ell$ and as a result FeSe is in the clean limit. Furthermore, the carrier density of this compensated metal, $n \sim2 \times 3.6 \times 10^{20}$\,cm$^{-3}$ \cite{Watson2015b}, is close to the density of paired electrons, $n_s$,
estimated from muon spin rotation (using a penetration depth of  $\lambda \sim 391$~nm) \cite{Biswas2018}
suggesting that all carriers have condensed at low temperatures.

\begin{figure}[h!]
	\centering
	\includegraphics[trim={0cm 0cm 0cm 0cm},width=0.90\linewidth,clip=true]{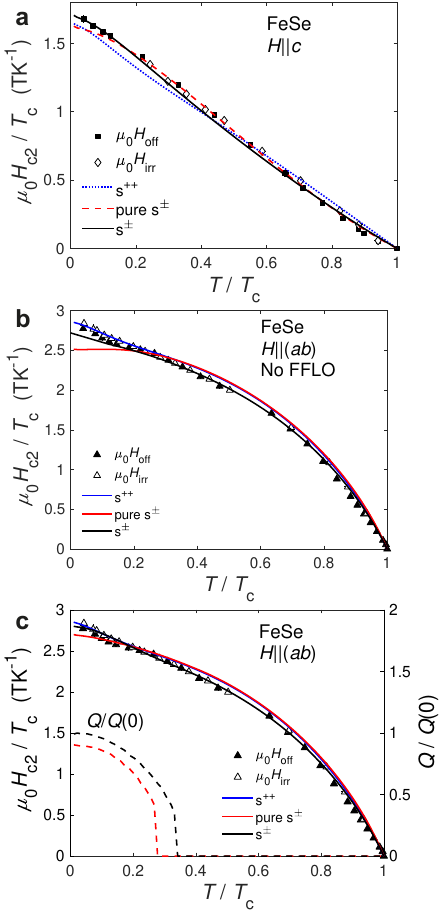}
	\caption{
			Upper critical fields of FeSe for (a) $H||c$ (squares) and (b) and (c) $H||(ab)$ (triangles).
			In (a) the dotted blue line corresponds to $s^{++}$ intraband pairing ($\lambda_{11}=0.81$, $\lambda_{22}=0.29$ and $\lambda_{12}=\lambda_{21}=0.1$),
the dashed red line $s^{\pm}$ pure interband pairing ($\lambda_{11}=\lambda_{22}=0$, $\lambda_{12}=\lambda_{21}=0.5$) and the solid black line interband $s^{\pm}$ pairing with a dominant band ($\lambda_{11}=0.81$, $\lambda_{22}=0$ and $\lambda_{12}=\lambda_{21}=0.5$).
			Here the $s^{\pm}$ interband model is best, with $\alpha=0$ and $\eta\sim0.01$.
			(b) Two-band models of the upper critical field when $H||(ab)$ for case where no FFLO state is present.
			The solid blue curve represents the $s^{++}$ case, the solid red curve the pure $s^{\pm}$ case and the solid black curve the $s^{\pm}$ case with a dominant band (same band coupling parameters as above).
			(c) Two-band models when $H||(ab)$ with the inclusion of an FFLO modulation.
			The curves represent the same cases as in (b), and the dashed lines of the same colour are the FFLO modulation, $Q/Q(0)$ (right-axis), for each case.
			Again, the $s^{\pm}$ case with a dominant band offers the best description using $\alpha=1.6$ and $\eta\sim0.02$.
	}
	\label{Fig3}
\end{figure}

To further assess the changes in the superconducting anisotropy at the lowest experimental temperature,
 we have performed an angle-dependent study of the upper critical field, $\mu_{0}H_{\rm{c2}}(\theta)$,
 for different samples of FeSe measured by transport and torque at $0.35\,$K, as presented in Fig.~\ref{Fig2}(c) (raw data shown
in Fig.~S1 in the SM).
The angular dependence of the upper critical field can normally be described by the anisotropic single-band Ginzburg-Landau (GL) theory \cite{Bennemann2008}.
However, we observe deviation from this  model, along with the 2D Tinkham model, while trying to describe transport and torque data of FeSe (Fig.~\ref{Fig2}(c)).
Instead,   the observed behaviour is described by accounting for the contribution of a second band to the angular dependence
that can be strongly temperature dependent and vary with the strength of square of Fermi velocities, as detailed in Ref.~\onlinecite{Gurevich2003}.

{\bf Two-band description.}
In order to describe the complete temperature dependence of $H_{\rm{c2}}(T)$ for FeSe in the two magnetic field orientations, a two-band  isotropic model
in the clean limit is considered, as detailed in Ref.~\onlinecite{Gurevich2010} and previously applied to CaKFe$_4$As$_4$ in Ref.~\cite{Bristow2020CaK}.
This model accounts for the presence of two different bands, with intraband ($\lambda_{11}$, $\lambda_{22}$) and interband scattering ($\lambda_{12}$, $\lambda_{21}$), and includes paramagnetic effects.
It allows for the presence of an FFLO inhomogeneous state at high-fields and low temperatures \cite{Gurevich2010}.
In many iron-based superconductors, the pairing is expected to be mediated by spin-fluctuations leading
to a sign changing  $s^{\pm}$ order parameter described by dominant interband coupling parameters with
$\lambda_{11}\lambda_{22}\ll\lambda_{12}\lambda_{21}$ (here the coupling constants are reported positive but
the sign of the products are unaffected even if the signs are negative and correspond to the expected repulsive interactions ) \cite{Hirschfeld2011, Cho2011,Tarantini2011}.
Extensive simulations have established that the shape of the upper critical field is strongly
sensitive to the values of $\eta=v_{2}^{2}/v_{1}^{2}$, which can change for
different field orientations due to the variation of velocities on the Fermi surface, as shown in Ref.~\onlinecite{Bristow2020CaK} and
Fig.~S4 in the SM.
We find that the temperature dependencies of $H_{\rm{c2}}(T)$ for FeSe in both orientations is best described by parameters
that consider interband pairing and a intraband scattering on a single band, using $\lambda_{11}=0.81$, $\lambda_{22}=0$ and $\lambda_{12}=\lambda_{21}=0.5$, as shown in Figs.~\ref{Fig3} and
Fig.~S6 in the SM.
For $H||c$ we use $\alpha=0$ and $\eta \sim0.01$, while for $H||(ab)$ $\alpha_{1}=1.6$, $\alpha_{2}=0$ (Pauli paramagnetic effects only on the dominant band) and $\eta\sim0.02$.
These $\eta$ values reflect a strong anisotropy of the Fermi velocity between the two bands.
Interestingly, the value of the Fermi velocity (expressed in $\hbar^{-1}$ units) for the dominant band for $H||c$
has $v_{\rm{1}}\sim315 \,\mathrm{meV\AA}$, which is remarkably similar to $\sim360\,\mathrm{meV\AA}$ associated with the hole band $\beta$ orbit in previous  ARPES studies \cite{Reiss2017,Watson2015c}.
The second band has a velocity of $v_{\rm{2}}\sim36\,\mathrm{meV\AA}$, that is
 much lower than that expected for the outer electron band  $\sim450\,\mathrm{meV\AA}$  \cite{Reiss2017,Watson2015c}.
Additionally, we can compare these values of velocities using the Pippard coherence length, $\xi_{0}$, that quantifies the size of Cooper pairs.
Using the coherence lengths of $\xi\sim4.6\,$nm above and $\xi=\hbar v_{\rm{F}}/\pi\Delta(0)$, we find that $v_{\rm{F}}$ varies
from $332$, $216$ and $56\,\mathrm{meV\AA}$, respectively (assuming the presence of the three different superconducting gaps in FeSe mentioned above).
Thus, based on these values one could suggest that the two-band model used here reveals the role played by the hole band and another band with a small gap and low velocity.
Alternatively, the extracted velocities could
 reflect the behaviour of a dominant band involved in pairing and the two values
of velocities would correspond to the in-plane and out-of-plane averaged components (see Fig.~\ref{Fig1}(f)).

The in-plane model for FeSe  ($H||(ab)$) uses the same coupling parameters a slightly larger $\eta$ value and lower velocities of $v_{\rm{1}}\sim140\,\mathrm{meV\AA}$, $v_{\rm{2}}\sim20\,\mathrm{meV\AA}$.
For this field orientation we use the coherence length of $\xi\sim2.8\,$nm along with the same values of $\Delta$ as above which gives predicted velocities of $132$ and $34\,\mathrm{meV\AA}$.
Additionally due to a finite Maki parameter of $\alpha\sim1.6$, the model allows for the stabilization of a FFLO state below $T_{\rm{FFLO}}\sim0.3T_{\rm{c}}\sim2.5\,$K as shown in Fig.~\ref{Fig3}(c).
The formation of the FFLO state in a system with a cylindrical Fermi surface requires a large Zeeman energy and a critical Maki's parameter of $\alpha_{\rm{c}}=4.76$ \cite{Song2019}, compared to  $\alpha_{\rm{c}}=1.8$ \cite{Gruenberg1966} for a 3D ellipsoidal Fermi surface.
A FFLO state can be realized in clean materials with weak scattering of quasiparticles.
It manifests as a change in slope in the upper critical field at low temperatures \cite{Gurevich2010} and it has been suggested to occur in FeSe \cite{Kasahara2020}.
Bands which could be involved
in this effect are expected to be
associated with shallow Fermi surface pockets with small Fermi energies
which can be spin polarized in large magnetic fields.
At low temperatures possible candidates are
the inner hole pocket
centered at the $Z$ point, which is pushed down below the Fermi level  inside the nematic electronic phase, or the electron pocket centered
at the zone corner which is lifted close to the Fermi level ($\sim3\,$meV),
as detected by ARPES \cite{Coldea2018},

Besides the model described above for FeSe,
 we have considered coupling  parameters that could be related
  to $s^{++}$ pairing in the presence of the orbital fluctuations and spin-orbit coupling, as suggested for LiFeAs \cite{Saito2015}.
By investigating different combinations of the band coupling parameters
 we find that intraband pairing with
 $\lambda_{11}=0.81$, $\lambda_{22}=0.29$ and $\lambda_{12}=\lambda_{21}=0.1$
 describe reasonably the temperature dependence of the upper critical field data in FeSe
 for both orientations in magnetic field.
  However, by assessing the best fit, it appears
  that interband scattering promoted by  spin fluctuations could dominate over intraband scattering
 (see Fig.~\ref{Fig3},
 Fig.~S6
 and
 Table~S1 in the SM).
 The $s^{\pm}$ coupling parameters also described
 well the upper critical field of thin flakes of  FeSe \cite{Farrar2020}
 (see Fig.~S7 in the SM)
 and Cu-FeSe  \cite{Zajicek2022},
 which are more sensitive to disorder and the degree of anisotropy is suppressed ($\eta \sim 0.2$).
As in other iron-based superconductors, the presence of several scattering channels in a multi-band system,
like FeSe, can increase the upper critical field far above the single-band limit, caused by the relative weight of different scattering channels.
On the other hand, the shallow bands in FeSe are likely to be involved in the stabilization of the FFLO state,
which is very fragile to disorder, as found for thin flakes of FeSe \cite{Farrar2020}
 (see Fig.~S7 in the SM)
 and Cu-FeSe  \cite{Zajicek2022}.
The temperature dependence of the upper critical field of FeSe reflects the behaviour of two dominant superconducting gaps
that either reside on different sheets of the cylindrical Fermi surface (one on a large hole band and another one on a shallow small band),
or the strong gap anisotropy of the hole band, which may have nodes or deep minima at the long axis of the ellipse \cite{Sprau2017}.
Furthermore, ARPES studies suggest  that there is significant anisotropy of the superconducting gap in all momentum directions
which needs to be taken into account \cite{Kushnirenko2018}.

The upper critical field values reported here are in good agreement with previous transport and torque
studies on FeSe \cite{Kasahara2020,Jong2020}.
All these studies reveal a linear temperature
dependence of the upper critical field for $H||c$,  as shown in
Fig.~S5 in the SM,
despite a small variation in the reported values of $T_{\rm{c}}$.
The precise definition of $T_{\rm c}$ for FeSe can vary slightly between
 thermodynamic and transport studies which can be caused by the growth process itself \cite{McQueen2009,Bohmer2016}
 and potentially any small applied strain if the crystals are glued during the experiments \cite{Ghini2021}.
On the other hand, the upper critical field boundaries  for  $H||(ab)$
show additional features at low temperature which differ
more between various studies (see Fig.~S5) \cite{Jong2020,Kasahara2016}.
 Furthermore, recent heat capacity suggests that sizable field-induced Gaussian superconducting fluctuations
could affect the precise upper boundaries of the upper critical field  \cite{Hardy2020}.

 There has been a series of proposed field-induced magnetic transitions in FeSe.
The non-saturating behaviour of $H_{\rm{c2}}$ for $H||c$ below $1.3\,$K was linked to unconventional pairing
due to the spin-splitting of the Fermi surface \cite{Kasahara2016,Watashige2017}.
Furthermore,  the thermal conductivity
has a transition field, $\mu_0 H^{\star}$,
that remains almost constant around $14\,$T, and  lower than the $\mu_0 H_{\rm{irr}}$ obtained from torque data \cite{Kasahara2014,Kasahara2016}.
Other studies  propose an
in-plane field-induced phase transition originating from the inherent spin-density-wave instability of
quasiparticles in FeSe \cite{Jong2020}.
Despite all these findings, the upper critical field of FeSe tuned by pressure or chemical pressure \cite{Bristow2019Hc2,Kim2022,Mukasa2023}
maintains an almost linear dependence for $H||c$, suggesting that multi-band effects are necessary to describe
the upper critical field of iron-chalcogenide superconductors.
Similarly, most of the temperature
dependence studies of the superfluid density and specific heat   \cite{Biswas2018,Hardy2019}
also confirmed the relevance of multi-band and anisotropic effects in FeSe.

\section{Conclusion}
In summary, we have experimentally mapped the temperature dependence of the upper critical fields  of FeSe
based on transport and torque measurements  up to $35\,$T and down to $0.35\,$K for two orientations in magnetic field.
Employing a two-band model in the clean limit, we estimate the band coupling constants, which suggest
more dominant interband scattering promoted by spin fluctuations in FeSe.
Additionally, for magnetic fields aligned in the conducting plane the temperature
dependence of the upper critical field is consistent with the emergence of an FFLO state at low temperatures.
Further theoretical studies need to take into account the exact details of the electronic structure as well as
 the gap anisotropy  to explain
 the specific features of the upper critical field as well as other relevant
 temperature-dependent superconducting quantities of FeSe.

In accordance with the EPSRC policy framework on research data, access to the data will be made available from ORA (https://ora.ox.ac.uk/).

\section{Acknowledgements}
We thank  P. Reiss for technical support and Andreas Kreisel and Peter Hirschfeld for useful discussions.
This work was partly supported by EPSRC (EP/I004475/1, EP/I017836/1) and the Oxford Centre for Applied Superconductivity.
DFT calculations were performed on the University of Oxford Advanced Research Computing Service \cite{ARC}.
AAH acknowledges the support of the Oxford Quantum Materials Platform Grant  (EPSRC, EP/M020517/1).
Part of this work was supported HFML-RU/FOM, member of the European Magnetic Field Laboratory (EMFL) and by 
EPSRC (UK) via its membership to the EMFL (EPSRC, EP/N01085X/1).
We also acknowledge financial support of the John Fell Fund of the Oxford University.
JCAP acknowledges the support of St Edmund Hall, University of Oxford, through the Cooksey Early Career Teaching and Research Fellowship.
AIC acknowledges an EPSRC Career Acceleration Fellowship (EP/I004475/1) and the Oxford Centre for Applied Superconductivity.
AIC is grateful to KITP for hospitality and
 thus this research was supported in part by the National Science Foundation 
 under Grants No. NSF PHY-1748958 and PHY-2309135.

\bibliography{FeSe_Hc2_may2020}

\newpage

\newcommand{\blue}{\textcolor{blue}}
\newcommand{\bdm}[1]{\mbox{\boldmath $#1$}}

\renewcommand{\thefigure}{S\arabic{figure}} 
\renewcommand{\thetable}{S\arabic{table}} 

\onecolumngrid

\newlength{\figwidth}
\figwidth=0.48\textwidth

\setcounter{figure}{0}

\newcommand{\fig}[3]
{
	\begin{figure}[!tb]
		\vspace*{-0.1cm}
		\[
		\includegraphics[width=\figwidth]{#1}
		\]
		\vskip -0.2cm
		\caption{\label{#2}
			\small#3
		}
\end{figure}}

 \begin{center}
{\bf Supplemental Materials for {\it Multi-band description of the upper critical field of bulk FeSe} }
 \end{center}

\vspace{3cm}

  \begin{figure*}[htbp]
  	\centering
  \includegraphics[trim={0cm 0cm 11cm 0cm}, width=0.47\linewidth,clip=true]{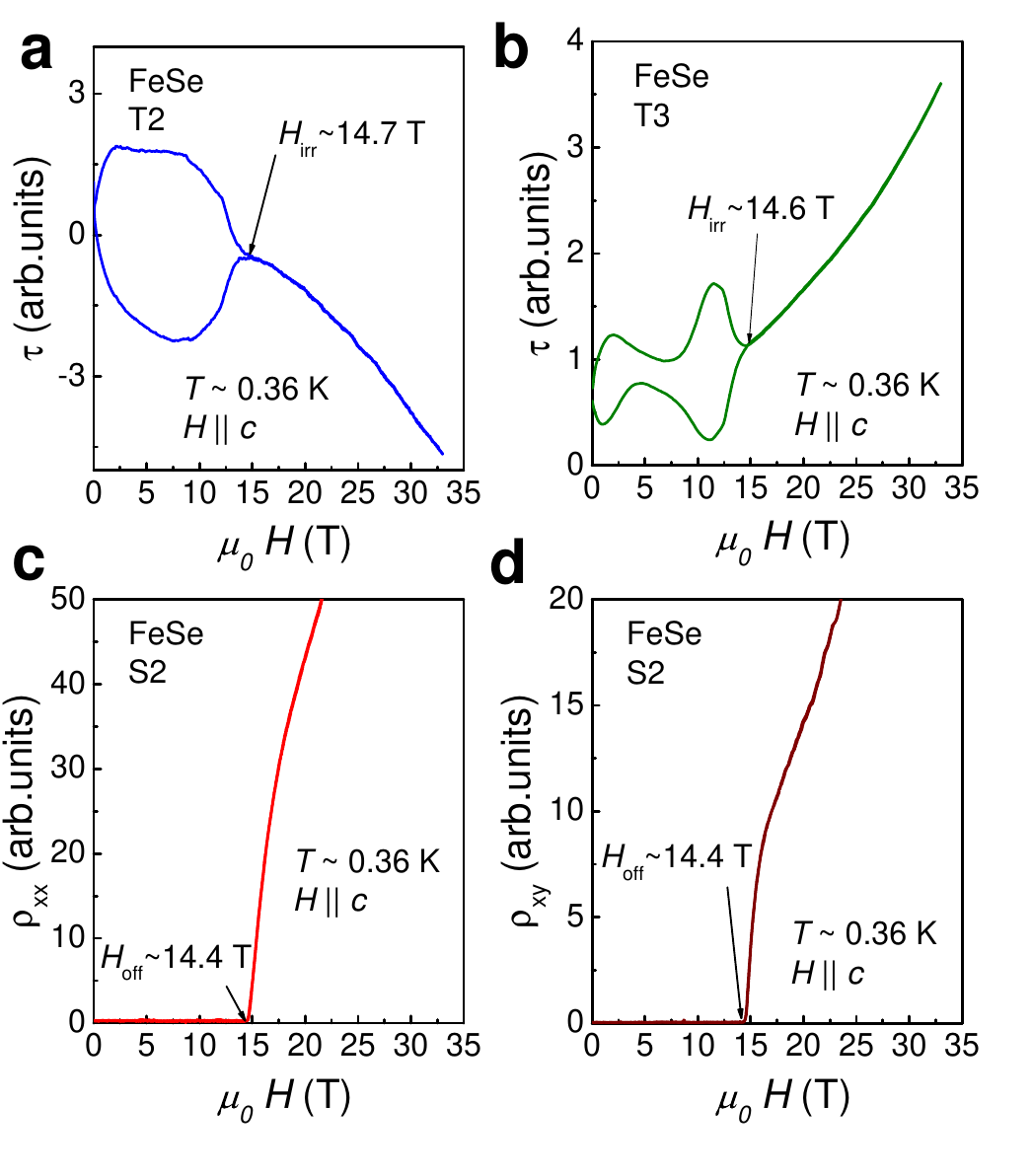}
 \includegraphics[trim={0cm 0cm 11cm 0cm}, width=0.47\linewidth,clip=true]{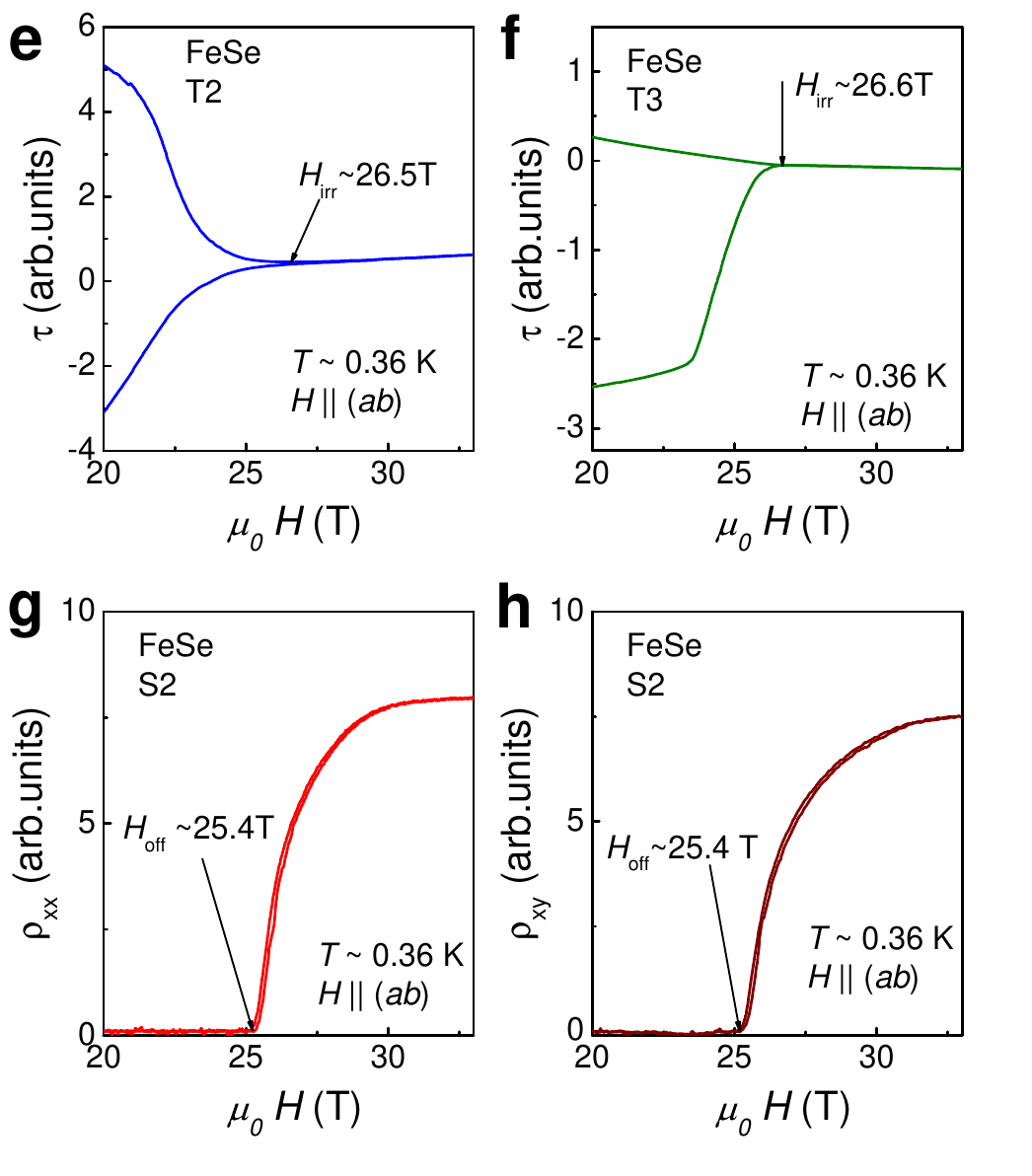}
 \caption{
\textbf{Additional transport and torque data of FeSe as a function of magnetic field for two different orientations.}
(a) and (b) Torque as a function of magnetic field for two samples (T2 and T3) of FeSe at base temperature with irreversibility field extraction for $H||c$.
(c) and (d) resistivity and Hall resistivity, respectively, against magnetic field for the sample FeSe sample (S2) with critical field extraction for $H||c$.
(e) and (f) Torque as a function of magnetic field for two samples (T2 and T3) of FeSe at base temperature with irreversibility field extraction for $H||(ab)$.
(g) and (h) resistivity and Hall resistivity, respectively, against magnetic field for the sample FeSe sample (S2) with critical field extraction for $H||(ab)$.}
\label{FigSM_rhoT}
\end{figure*}

\begin{figure*}[htbp]
\centering
       	\includegraphics[trim={0cm 0cm 0cm 0cm}, width=1\linewidth,clip=true]{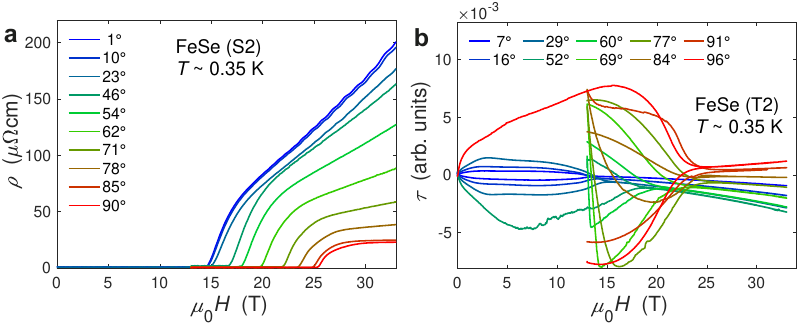}
 \caption{{\bf The angular dependent studies of upper critical field in FeSe.}
 	The field dependence of (a) resistivity and (b) torque for two different single crystals
 rotated in magnetic field from $H||c$ ($\theta=0^{\circ}$) towards $H||({\rm ab})$ ($\theta=90^{\circ}$).
 }
 \label{figSM:Rotation_RawData}
\end{figure*}

 \begin{figure*}[htbp]
	\centering
	\includegraphics[width=0.85\linewidth,clip=true]{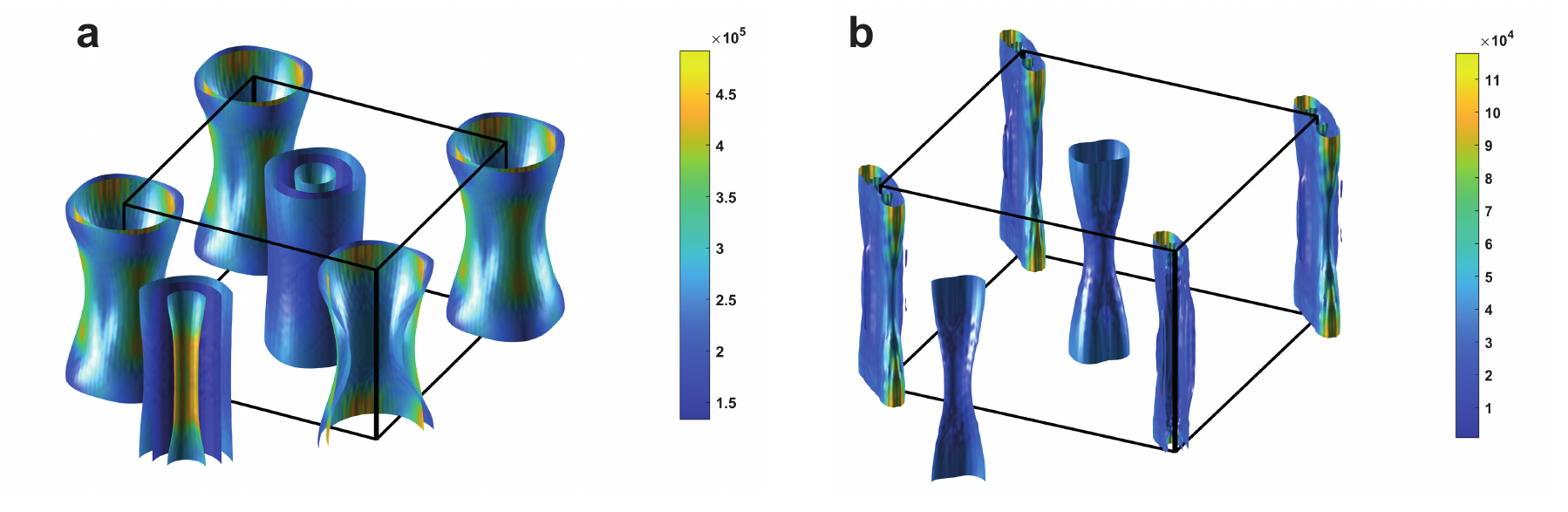}
	\caption{\textbf{Fermi surface of FeSe.}
		(a) Fermi surface with velocity colour of the orthorhombic Fermi surface and (b) the shifted Fermi surface
to match quantum oscillations data at low temperatures \cite{Watson2015a,Terashima2014,Coldea2019}.
		Solid lines indicate the Brillouin zone.
				The calculations were performed using Wien2k and GGA approximation and the experimental lattice parameters
described by the  $Cmma$ symmetry group were
$a=5.3082\,$\AA, $b=5.3346\,$\AA, $c=5.4864\,$\AA ~and $z_{\rm{Se}}=0.2357$.
		The calculated value of anisotropy, $\Gamma$=$\lambda_{c}/\lambda_{ab}$  (obtained
using a similar approach to Ref.~\onlinecite{Hashimoto2010}) is 4.1
for the unshifted Fermi surface and 4.6 for shifted Fermi surface.
	}
	\label{FigSM:FS}
\end{figure*}

\clearpage
\newpage

\vspace{0.5cm}
{\bf The upper critical fields using a two-band model.}

To describe the entire temperature dependence of $H_{\rm{c2}}(T)$ of bulk FeSe for both orientations,
we use a two-band model in the clean limit. This model includes paramagnetic effects and
allows the emergence of a FFLO state, as detailed in Ref.~\onlinecite{Gurevich2010}.
The values of the upper critical field for different temperatures (in units of reduced temperature $t=T/T_{\rm c}$)
are found from estimating the following expressions:

\begin{multline}
a_{1}[\ln t + U_{\rm{1}}]+a_{2}[\ln t + U_{\rm{2}}]
+ [\ln t + U_{\rm{1}}][\ln t + U_{\rm{2}}] = 0, \;\;\;\;
\label{TwoBandFFLO_equ}
\end{multline}

where $a_{\rm{1}}=(\lambda_{\rm{0}}+\lambda_{\rm{-}})/2w$,
$a_{\rm{2}}=(\lambda_{\rm{0}}-\lambda_{\rm{-}})/2w$,
$\lambda_{\pm}=\lambda_{11}\pm\lambda_{22}$,
$w=\lambda_{11}\lambda_{22}-\lambda_{12}\lambda_{21}$ and
$\lambda_{\rm{0}}=(\lambda_{-}^{2}+4\lambda_{12}\lambda_{21})^{\rm{1/2}}$.
Here $\lambda_{ij}$ represents a coupling constant between bands $i$ and $j$.
As the cross-terms $\lambda_{12}$ and $\lambda_{21}$ only appear together we set $\lambda_{21}=\lambda_{12}$ for simplicity.

The values of the coupling constants allows us to establish the dominant pairing
corresponding either to $s^{++}$ pairing when the parameter $w>0$
or to the $s^{\pm}$ pairing for $w<0$.

$U_1$ and $U_2$ are defined as follows:


\begin{multline}
U_1 = 2e^{q^{2}}\rm{Re}\it \sum_{n=0}^{\infty}\int_{q}^{\infty} du \cdot {}e^{-u^{2}}
\Bigg( \frac{u}{n+1/2} - \frac{t}{\sqrt{b}}\rm{tan^{-1}}\Bigg[ \it \frac{u\sqrt{b}}{t(n+\rm{1/2}) + \it i\alpha{}b} \Bigg] \Bigg)
\label{TwoBandFFLO_U1_equ}
\end{multline}

\begin{multline}
U_2 = 2e^{q^{2}s}\rm{Re}\it \sum_{n=0}^{\infty}\int_{q\sqrt{s}}^{\infty}du \cdot {}e^{-u^{2}}
\Bigg( \frac{u}{n+1/2} - \frac{t}{\sqrt{b\eta}}\rm{tan^{-1}}\Bigg[ \it \frac{u\sqrt{b\eta}}{t(n+\rm{1/2}) + \it i\alpha{}b} \Bigg] \Bigg)
\label{TwoBandFFLO_U2_equ}
\end{multline}

The variables used for this two-band model, $b$, $\alpha$, $q$, $\eta$ and $s$, are defined as

\begin{multline}
\;\;\;\;\;\;\;\;\;\;\;\;\;\;
b = \frac{\hbar^2 v_{1}^{2}H_{\rm c2}}{8\pi\phi_{\rm{0}}{k_{\rm{B}}}^2\it{T}_{\rm{c}}^{2}}, \;\;\;\;
\alpha = \frac{4\mu\phi_{\rm{0}} {k_{\rm{B}}}\it{T}_{\rm{c}}}{\hbar{}^2 v_{1}^{2}},
q^{2} = \frac{Q^2 \phi_{\rm{0}}\epsilon_{1}}{2\pi H_{\rm c2}}, \;\;\;\;
\eta = \frac{v_{2}^{2}}{v_{1}^{2}}, \;\;\;\;
s = \frac{\epsilon_{2}}{\epsilon_{1}}.
\;\;\;\;\;\;\;\;\;\;\;\;
\label{TwoBandFFLO_Params_equ}
\end{multline}

where

$v_{i}$ is the in-plane Fermi velocity of band $i$,

$\epsilon_{i}$ is the mass anisotropy ratio,

$\epsilon_{i}= m^{\perp}_{i}/m^{\parallel}_{i}$ is related to the ratio between the gradients near $T_{\rm{c}}$ in different field orientations,

$s$ is band mass anisotropy between the two bands and $ \gamma= \epsilon^{-1/2}$.

$Q$ is the magnitude of the FFLO modulation
and it is found for a given temperature when $H_{\rm{c2}}$ is maximal ($dH_{\rm{c2}}/dQ=0$).
We use the scaled value at zero temperature, $Q/Q(0)$, to enable the comparison with
the upper critical field in the same graphs.

\begin{figure*}[htbp]
	\centering
	\includegraphics[width=1.0\linewidth,clip=true]{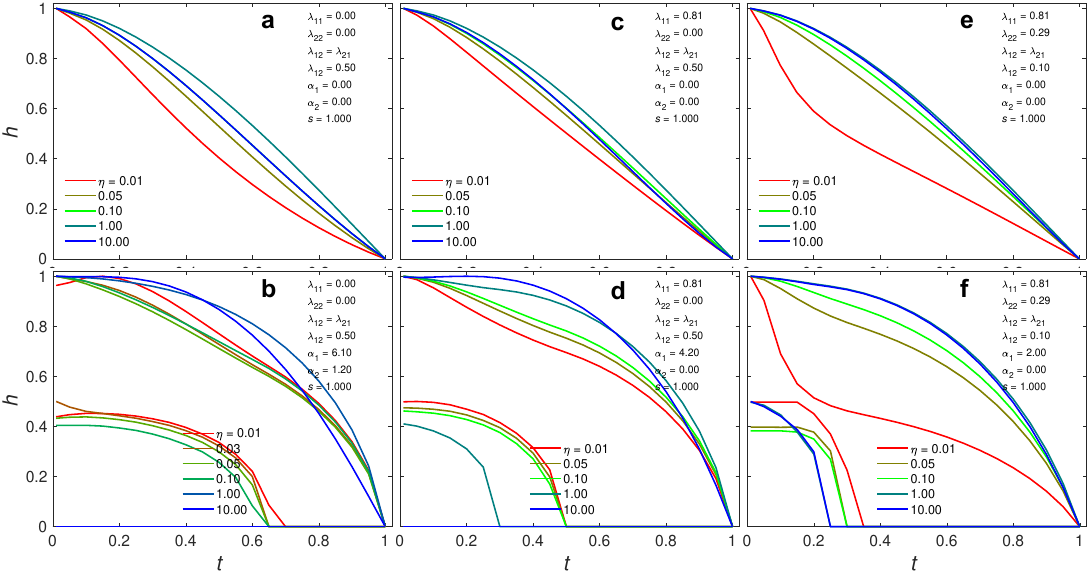}
	\caption{\textbf{Two-band upper critical field simulations for different band coupling parameters and values of $\eta$.}
		Simulations of the reduced upper critical field $h$ using the two-band model previously discussed in detail in Ref.~\onlinecite{Bristow2020CaK}.
		The parameters listed in each panel correspond to the $s^{\pm}$ pairing in (a) and (b) with $\lambda_{11}=\lambda_{22}=0$, $\lambda_{12}=\lambda_{21}=0.5$,
		$s^{\pm}$ pairing in (c) and (d) with $\lambda_{11}=\lambda_{22}=0.81$, $\lambda_{12}=\lambda_{21}=0.5$,
		and the $s^{++}$ pairing in (e) and (f) with $\lambda_{11}=0.81$, $\lambda_{22}=0.29$ and $\lambda_{12}=\lambda_{21}=0.1$, $\alpha_{1}=0.95$ for $H||c$.
		For panels (b), (d) and (f) the lower curves (left corners) represent the normalized magnitude of the FFLO modulation to the zero-temperature
value, $Q/(2Q(0))$.}
	\label{FigSM_TwoBand_Simulations}
\end{figure*}

Various simulations of upper critical fields using this two-band model are shown in Fig.~\ref{FigSM_TwoBand_Simulations}.
The simulations of $H_{\rm c2}$ for $H||(ab)$  assumed that the bands have the same anisotropy
parameter $\epsilon_{1} = \epsilon_{2} = \epsilon$.
Other parameters are estimated based on the fitted values using a single-band model
and the velocities are minimised locally to find an optimum solution.

A rescaling of the upper critical field has been performed previously in Ref.~\onlinecite{Gurevich2010}.
by mapping $H_{\rm{c2}}^{c}\rightarrow H_{\rm{c2}}^{ab}$ using
$q_{ab} \rightarrow q_{c} \epsilon^{-3/4}$,
$\alpha_{ab} \rightarrow \alpha_{c}  \epsilon^{-1/2}$,
$b^{1/2}_{ab} \rightarrow b^{1/2}_{c} \epsilon^{1/4}$ in $U_1$,
$(\eta_{ab} b_{ab})^{1/2} \rightarrow (\eta_{c} b_{c}^{1/2})\epsilon^{1/4}$ in $U_2$.
The values of the $\alpha$ parameters change as $\alpha_{ab,1}\rightarrow\alpha_{c} \epsilon^{-1/2}$, and $\alpha_{ab,2}\rightarrow\alpha/(\eta\epsilon^{1/2})$.
However, this re-scaling is not employed here as the ratio between the Fermi velocities $\eta$ is a variable parameter
that can be changed between the two field orientations.

\begin{table*}[htbp]
	\caption{
		Two-band parameters for models of the upper critical fields in FeSe in different pairing limits by assuming different band coupling constants.
	}
	\label{Table:FeSe_TwoBand}
	\begin{center}
			\begin{tabular}{c||cc|cc|cc||cc|cc|cc}
				\hline
				\hline
				FeSe & \multicolumn{6}{|c||}{$\mu_{0}H_{\rm{off}}$ (S1)} & \multicolumn{6}{|c}{$\mu_{0}H_{\rm{irr}}$ (T1)} \\
				& \multicolumn{2}{|c|}{$s^{\pm}$ dominant} & \multicolumn{2}{|c|}{pure $s^{\pm}$} & \multicolumn{2}{|c||}{$s^{++}$} & \multicolumn{2}{|c|}{$s^{\pm}$ dominant} & \multicolumn{2}{|c|}{pure $s^{\pm}$} &
\multicolumn{2}{|c}{$s^{++}$} \\
				$H||$ & $c$ & $(ab)$ & $c$ & $(ab)$  & $c$ & $(ab)$  & $c$ & $(ab)$  & $c$ & $(ab)$  & $c$ & $(ab)$  \\
				\hline
				$T_{\rm{c}}$ (K) & 9.08 & 9.08 & 9.08 & 9.08 & 9.08 & 9.08 &
				8.5 & 8.5 & 8.5 & 8.5 & 8.5 & 8.5 \\
				$\lambda_{11}$ & 0.81 & 0.81 & 0 & 0 & 0.81 & 0.81 &
				0.81 & 0.81 & 0 & 0 & 0.81 & 0.81 \\
				$\lambda_{22}$ & 0 & 0 & 0 & 0 & 0.29 & 0.29 &
				0 & 0 & 0 & 0 & 0.29 & 0.29 \\
				$\lambda_{12}$ & 0.5 & 0.5 & 0.5 & 0.5 & 0.1 & 0.1 &
				0.5 & 0.5 & 0.5 & 0.5 & 0.1 & 0.1  \\
				$\lambda_{21}$ & 0.5 & 0.5 & 0.5 & 0.5 & 0.1 & 0.1 &
				0.5 & 0.5 & 0.5 & 0.5 & 0.1 & 0.1 \\
				$\alpha_{1}$ & 0 & 1.6 & 0 & 1.6 & 0 & 1.6 &
				0 & 1.6 & 0 & 1.6 & 0 & 1.6 \\
				$\alpha_{2}$ & 0 & 0 & 0 & 0.5 & 0 & 0 &
				0 & 0 & 0 & 0.5 & 0 & 0 \\
				$v_{1}$ ($\mathrm{meV\AA}$)  & 315 & 137 & 438 & 153 & 277 & 124 &
				310 & 132 & 392 & 151 & 254 & 117 \\
				$v_{2}$ ($\mathrm{meV\AA}$)  & 39 & 19 & 98 & 68 & 48 & 39 &
				38 & 19 & 111 & 43 & 40 & 33 \\
				$\eta$ & 0.01 & 0.02 & 0.05 & 0.2 & 0.03 & 0.1 &
				0.01 & 0.02 & 0.08 & 0.08 & 0.025 & 0.08 \\
				$T_{\rm{FFLO}}$ (K) & -- & 2.8 & -- & 2.2 & -- & -- &
				-- & 2.63 & -- & 2.8 & -- & -- \\
				MSE ($10^{-2}$) & 1.37 & 1.18 & 1.54 & 1.70 & 7.63 & 1.51 &
				0.14 & 0.011 & 0.62 & 0.86 & 0.59 & 0.57 \\
							\hline
				\hline
		\end{tabular}
	\end{center}
\end{table*}

\begin{figure*}[htbp]
	\centering
	\includegraphics[width=0.5\linewidth,clip=true]{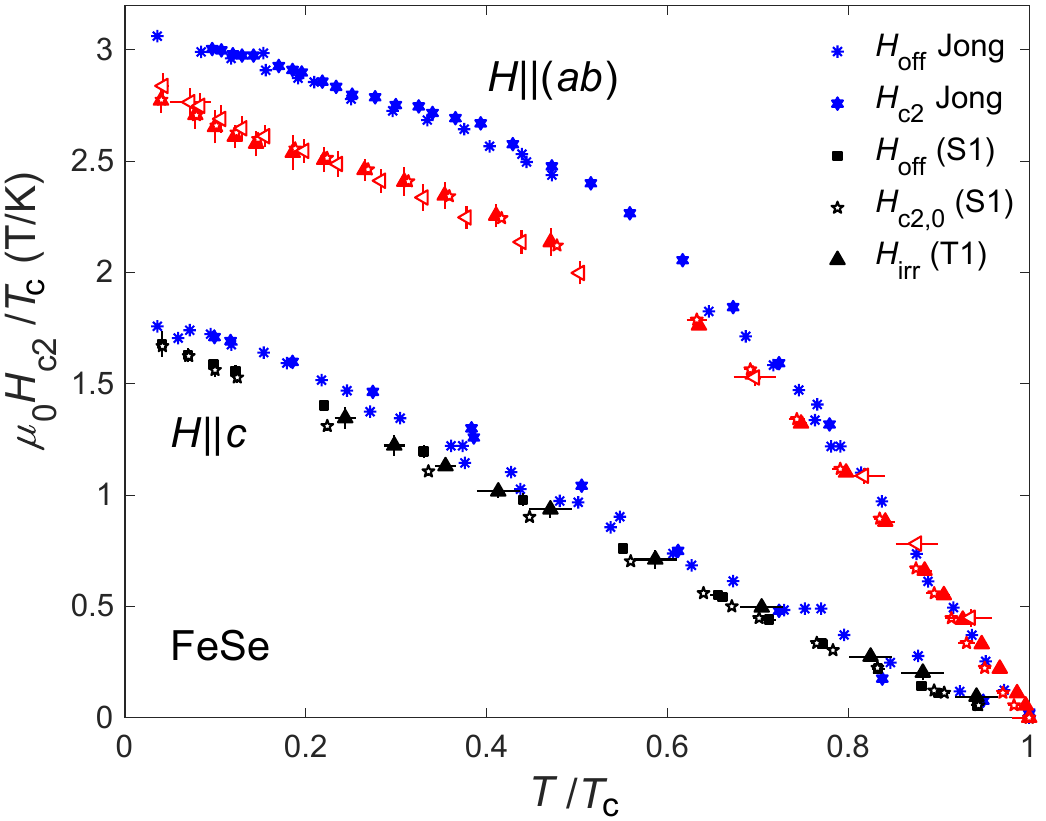}	\includegraphics[width=0.5\linewidth,clip=true]{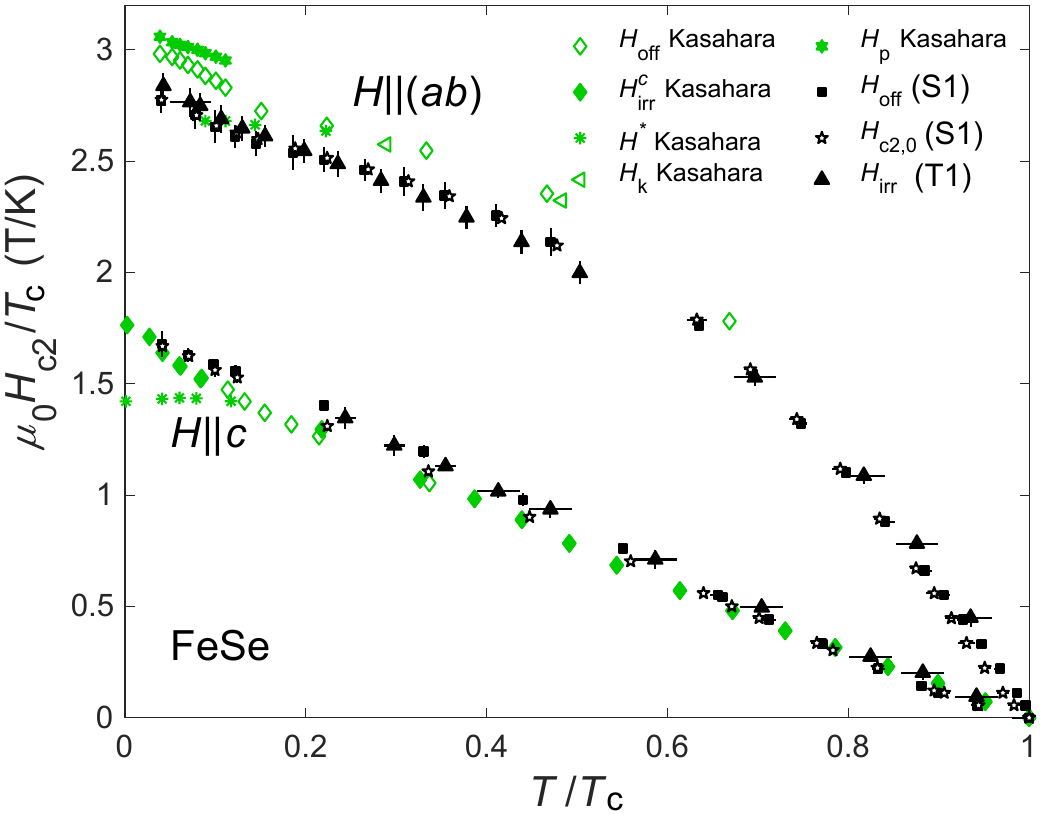}
	\caption{
		\textbf{Upper critical field comparisons for FeSe.}
		The two panels here compare the upper critical fields in reduced, $\mu_{0}H_{\rm{c2}}/T_{\rm{c}}$, against reduced temperature, $T/T_{\rm{c}}$.
		The panels compare our data presented in the main body of the paper against data from Jong \cite{Jong2020} and Kasahara \cite{Kasahara2020}.
	}
	\label{FigSM_FeSe_Hc2_Comparison}
\end{figure*}

\begin{figure*}[htbp]
	\centering
\includegraphics[trim={0cm 0cm 0cm 0cm}, width=0.90\linewidth,clip=true]{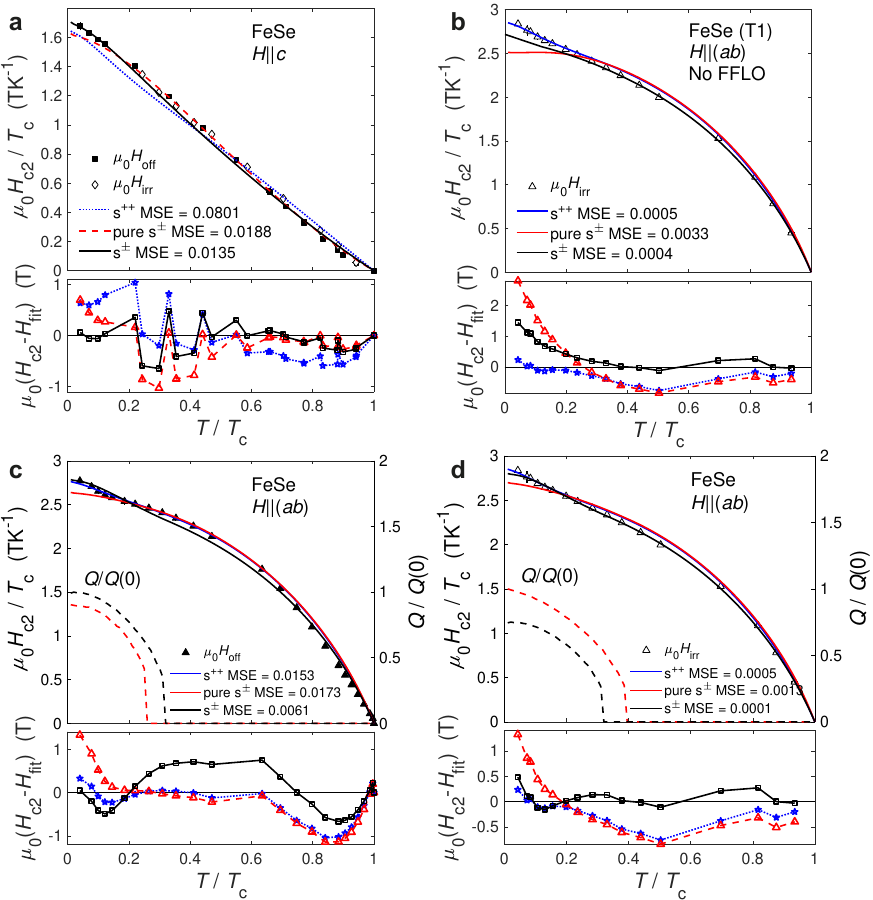}
	\caption{
		\textbf{Comparison of two-band fitting of the upper critical field in different limits for FeSe, with errors.}
We have fitted three different models applicable to both resistivity and torque data when $H||c$ in (a),
to torque data when $H||(ab)$ with no FFLO state in (b), to resistivity data when $H||(ab)$ with an FFLO modulation included in (c)
 and to torque data when $H||(ab)$ with an FFLO modulation included in (d).
	}
	\label{FigSM_FeSe_TwoBandModels}
\end{figure*}

\begin{figure*}[htbp]
\centering
       	\includegraphics[trim={0cm 0cm 0cm 0cm}, width=0.38\linewidth,clip=true]{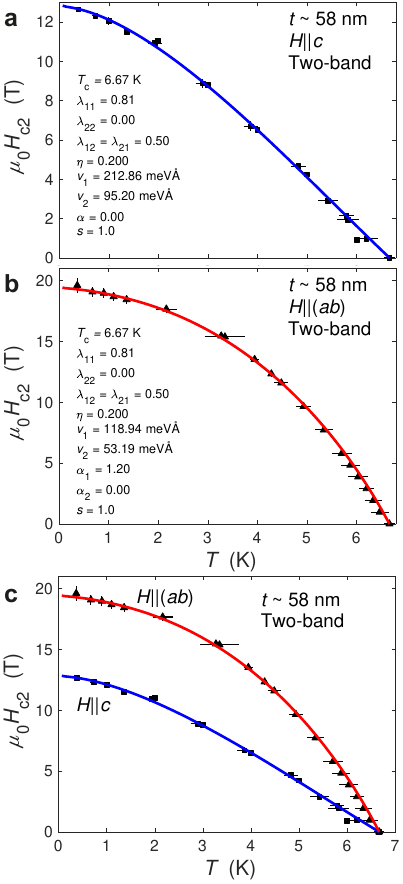}
 \caption{{\bf Upper critical field of a thin flake of FeSe.}
 	(a) Upper critical field of a 58~nm thin flake of FeSe fitted using a two-band model, as detailed in Ref.~\cite{Gurevich2010} for different orientations.
 	The fitting parameters are listed for $H||c$ in (a) and for $H||(ab)$ in (b) and are consistent with $s_{\pm}$ pairing in FeSe.
 	(c) The two magnetic field orientations are combined.
 }
 \label{figSM:thin_flakes}
\end{figure*}

\end{document}